\documentclass[%
 reprint,
 amsmath,amssymb,
 aps,
 pra,
]{revtex4-2}

\usepackage{gensymb}
\usepackage{braket}
\usepackage[normalem]{ulem}
\usepackage{graphicx}% Include figure files
\usepackage{dcolumn}% Align table columns on decimal point
\usepackage{bm}% bold math
\usepackage{upgreek}

\usepackage{soul}
\usepackage[export]{adjustbox}
\usepackage{xcolor}

\newcommand\thefont{\expandafter\string\the\font}
\usepackage{hyperref}

\begin{document}

\preprint{APS/123-QED}

\title{Measurement of the tune-out wavelength for $^{133}$Cs at $880$ nm}
\author{Apichayaporn Ratkata$^{1}$}
\author{Philip D. Gregory$^{1}$}
\author{Andrew D. Innes$^{1}$}
\author{Jonas A. Matthies$^{1}$}
\author{Lewis A. McArd$^{1}$}
\author{Jonathan M. Mortlock$^{1}$}
\author{M. S. Safronova$^{2}$}
\author{Sarah L. Bromley$^{1}$}
\author{Simon L. Cornish$^{1}$}%%
 \email{s.l.cornish@durham.ac.uk}
\address{
\mbox{$^{1}$Joint Quantum Centre (JQC) Durham-Newcastle, Department of Physics,} \mbox{Durham University, Durham, United Kingdom, DH1 3LE.} \mbox{$^2$Department of Physics and Astronomy, University of Delaware, Newark, Delaware 19716, USA.}  }

\date{\today}% It is always \today, today,

\begin{abstract}
We perform a measurement of the tune-out wavelength, $\lambda_{0}$, between the $D_{1}$, $6^2S_{1/2}\rightarrow6^2P_{1/2}$, and $D_{2}$, $6^2S_{1/2}\rightarrow6^2P_{3/2}$, transitions for $^{133}$Cs in the ground hyperfine state $(F=3, m_{F}=+3)$.  At $\lambda_{0}$, the frequency-dependent scalar polarizability is zero leading to a zero scalar ac Stark shift.  We measure the polarizability as a function of wavelength using Kapitza-Dirac scattering of a $^{133}$Cs Bose-Einstein condensate in a one-dimensional optical lattice, and determine the tune-out wavelength to be $\lambda_{0}=880.21790(40)_{\textrm{stat}}(8)_{\textrm{sys}}$ nm.  From this measurement we determine the ratio of reduced matrix elements to be $|\!\braket{6P_{3/2}\| d\|6S_{1/2}}\!|^{2}/|\!\braket{6P_{1/2}\| d\|6S_{1/2}}\!|^{2}=1.9808(2)$. This represents an improvement of a factor of 10 over previous results derived from excited-state lifetime measurements. We use the present measurement as a benchmark test of high-precision theory.
\end{abstract}

\maketitle

%\tableofcontents
%\section{\thefont}
%\thefont
\section{Introduction}\label{Intro}
Optical trapping is widely employed in experiments involving ultracold neutral atoms and molecules \cite{Grimm2000}. Optical fields can be engineered on the scale of the optical wavelength to produce various trapping geometries, including lattices \cite{Bloch2005}, ring traps \cite{Ramanathan2011,Moulder2012}, box potentials \cite{Gaunt2013,Mukherjee2017, Bause2021} and arrays of individual micro-traps \cite{Ebadi2020,Scholl2020}.  This, combined with the ability to confine any polarizable species, has resulted in numerous advances in metrology \cite{Ludlow2015}, control of single atoms and molecules \cite{Schlosser2001,Norcia2018,Cooper2018,Scholl2020,Ebadi2020,Liu2018,Anderegg2019}, and quantum simulation of interacting many-body systems \cite{Bloch2008,Lewenstein2012,Bloch2012,Gross2017}.  Refined optical trapping techniques can also lead to  exciting developments that will underpin future quantum technologies \cite{Amico2020,Awschalom2021,Altman2021}.

The dipole force experienced by atoms in an optical trap is proportional to the dynamic polarizability.  The polarizability varies with wavelength exhibiting poles, whenever the applied optical field matches a transition.  This wavelength dependence gives additional control over the optical potential where, for ground state atoms, optical frequencies red detuned of a transition give rise to attractive optical potentials and those frequencies that are blue detuned give rise to repulsive optical potentials. The poles in the polarizability lead to wavelengths between transitions where the polarizability is zero, commonly referred to as tune-out wavelengths \cite{Leblanc2007, Arora2011} or magic-zero wavelengths \cite{Herold2012}.  Precise knowledge of the polarizability is important for a number of applications including optical lattice clocks, quantifying lattice potentials, and as benchmarks for testing theoretical methods of calculating polarizability for more complex atoms such as Er and Dy \cite{Herold2012}.  Measurements of tune-out wavelengths are important as they allow the determination of multiple atomic properties including transition dipole matrix elements, oscillator strengths, and state lifetimes \cite{Herold2012,Trubko2017}.  Transition dipole matrix elements are fundamental properties of atoms as well as being crucial parameters for determining, for example, the blackbody radiation shift of atoms which is often a limiting systematic uncertainty in atomic clocks \cite{Nicholson2015}. A number of discrepancies between experimental results and between theory and experiment have been pointed out in the literature \cite{Meir2020,Arnold2019,Arnold2020,Heinz2020} recently, giving particular importance to further benchmark tests.

Tune-out wavelengths can also be used to create species-specific and state-specific optical trapping potentials \cite{Leblanc2007,Arora2011}.  Species-specific traps occur due to different atomic species having different transition wavelengths.  For different atomic species, the poles in the polarizability therefore occur at different wavelengths leading to different trapping potentials.  Species-specific trapping is useful in multi-species experiments and has allowed for studies of scattering in mixed dimensions \cite{Lamporesi2010} and the transfer of entropy between different atomic species to demonstrate novel cooling schemes \cite{Catani2009}. 
Within the same atomic species, atoms in different electronic states will experience different trapping potentials due to the different transition frequencies from the different states.  Even within the same electronic state it is possible to engineer state-specific potentials as the polarizability also depends on the polarization of the light interacting with the atom and the orientation of the atomic spin.  The light polarization and atomic spin will determine the transitions that are allowed by selection rules and hence make polarizability depend on both the total electronic angular momentum, $F$, and its projection, $m_{F}$.  In general, the polarizability is therefore composed into scalar, vector, and tensor polarizabilities where the scalar polarizability is the polarizability when averaging over all $m_{F}$ levels.  State-specific potentials can be used to engineer multi-particle entanglement \cite{Mandel2003}, spatiotemporal control of intraspecies interactions \cite{Clark2015}, and state-selective manipulation of quantum states \cite{Wang2015, Heinz2020}.    

Tune-out wavelengths have been experimentally measured both directly and indirectly.  Direct measurements are made by performing polarizability measurements around the tune-out wavelength, with experimental techniques including atom diffraction \cite{Herold2012}, parameteric heating \cite{Heinz2020}, and atom interferometry \cite{Leonard2015}.  Indirect measurements can be made by inferring tune-out wavelengths from measurements of state lifetimes, but can be limited by knowledge of branching ratios \cite{Herold2012}.  Previous experiments have directly measured tune-out wavelengths, using linearly polarized light, for different alkali-metal atoms including Li \cite{Copenhaver2019, Decamps2020}, K \cite{Holmgren2012,Trubko2017}, and Rb \cite{Catani2009,Herold2012,Leonard2015,Schmidt2016}, as well as for other atomic species including He \cite{Henson2015}, Sr \cite{Heinz2020}, and Dy \cite{Kao2017} and also for ground state NaK molecules \cite{Bause2020}.  However, despite many theoretical studies of $^{133}$Cs polarizability \cite{LeKien2013,Yu2016,Safronova2016,Jiang2020a}, so far no measurements of $^{133}$Cs scalar tune-out wavelength have been performed.  And yet $^{133}$Cs atoms are used in a wide range of applications including the definition of the second \cite{Essen1955}, the search for variations in fundamental constants, and tests of the standard model.  

In this paper, we report the experimental measurement of the scalar tune-out wavelength $\lambda_0$ of $^{133}$Cs atoms, hereafter just denoted Cs, in the ground hyperfine state at $\lambda_{0}\approx~880$~nm, between the $D_{1}$, $6^2S_{1/2}\rightarrow6^2P_{1/2}$, and $D_{2}$, $6^2S_{1/2}\rightarrow6^2P_{3/2}$, transitions. From this measurement we determine the ratio of reduced dipole matrix elements $|\!\braket{6P_{3/2}\| d\|6S_{1/2}}\!|^{2}/|\!\braket{6P_{1/2}\| d\|6S_{1/2}}\!|^{2}=1.9808(2)$.  This ratio is in agreement with previous results from lifetime measurements \cite{Patterson2015}, but with an error bar reduced by more than a factor of 10. Experimental determination of this ratio is also of particular interest  due to discrepancy between theoretical and experimental values in Ba$^+$ discussed in \cite{Arnold2019,Arnold2020}. We carry out a benchmark comparison with theoretical calculations of the Cs ratio, testing the methodology for determining theory uncertainties.

This paper is structured as follows.  In Sec. \ref{sec:Theory} we discuss the theoretical calculations of polarizability including its decomposition into scalar, vector, and tensor components.  We also explain how measurements of polarizability can be made using Kapitza-Dirac scattering that results from applying a pulsed optical lattice potential to the atoms.  In Sec. \ref{sec:Apparatus} we give a brief overview of the experimental apparatus and the production of Cs Bose-Einstein condensates (BECs).  In Sec. \ref{sec:Measurements} we discuss the lattice setup used to measure $\lambda_{0}$ and present the results.  In Sec. \ref{sec:Analysis} we discuss how we extract the scalar tune-out wavelength from our measurements.  We also discuss how the ratio of reduced matrix elements is extracted, and present theoretical calculations.  In Sec. \ref{sec:Outlook} we summarize the results  and give an outlook to future work.

\section{Theory}\label{sec:Theory}
\subsection{Polarizability}
Our experiments are performed using Cs atoms prepared in  $\ket{F=3, m_{F}=+3}$ in a magnetic field of 23.4(1)~G.  We calculate the polarizability including hyperfine structure following the methods described in detail elsewhere \cite{Safronova2006,LeKien2013, Jiang2020a}.  Below we summarize the main results.

The quantum state, $\ket{i}$, of an alkali-metal atom can be defined in terms of the quantum numbers $\ket{i}=\ket{\gamma,F,m_{F}}\equiv\ket{F,m_{F}}$.  $\mathbf{F}=\mathbf{I}+\mathbf{J}$, with $\mathbf{I}$ the nuclear spin, and $\mathbf{J}$ the electronic angular momentum.  For Cs, the nuclear spin $I=7/2$.  $\gamma$ represents the other quantum numbers used to define the state but we will drop $\gamma$ from the notation for simplicity. 

For an alkali-metal atom in $\ket{i}$ interacting with light of wavelength $\lambda$, with associated angular frequency, $\omega$, the general form of the frequency dependent polarizability can be decomposed as \cite{Leonard2015}
\begin{equation}
\begin{split}
\alpha_{i}(\omega)&= \alpha^{(0)}_{i}(\omega)-\xi\hat{\mathbf{k}}\cdot\hat{\mathbf{B}} \frac{m_{F}}{2F}\alpha^{(1)}_{i}(\omega)\\
&\qquad
+\left[\frac{3(\hat{\boldsymbol{\epsilon}}\cdot\hat{\mathbf{B}})^{2}-1}{2}\right]\frac{3m_{F}^{2}-F(F+1)}{F(2F-1)}\alpha^{(2)}_{i} (\omega).
\end{split}\label{eq:svtPolarizability}
\end{equation}

\noindent Here $\alpha^{K}_{i}$ are the scalar ($K=0$), vector ($K=1$), and tensor ($K=2$) components of the polarizability, $\xi=(I_{+}+I_{-})/I_{0}$ is the fourth Stokes parameter \cite{Budker2010} and quantifies the degree of circularity of the polarization with $I_{\pm}$ being the intensities of the different circular components and $I_{0}$ is the total intensity, $\hat{\mathbf{B}}$ is a unit vector in the direction of the magnetic field, and $\hat{\boldsymbol{\epsilon}}$ and $\hat{\mathbf{k}}$ are unit vectors in the direction of the polarization vector and wavevector, respectively, of the light interacting with the atoms.

This scalar polarizability can be further decomposed into \cite{Safronova1999}
\begin{equation}\label{eq:TotalScalar}
\alpha^{(0)}_{i}(\omega)=\alpha_{\textrm{core}}+\alpha_{\textrm{vc}}+\alpha^{(0)}_{\textrm{v},i}(\omega),
\end{equation}
\noindent  where there is a contribution to the polarizability from the core electrons, $\alpha_{\textrm{core}}$, a core modification due to the valence electron, $\alpha_{\textrm{vc}}$, and a contribution from the valence electron $\alpha_{\textrm{v},i}^{(0)}(\omega)$.  The excitation frequencies of the core electrons are far detuned from the laser frequencies considered here and so $\alpha_{\textrm{core}}$ and $\alpha_{\textrm{vc}}$ are treated as frequency independent. 
Calculations in the random-phase approximation (RPA) yield for Cs, $\alpha_{\textrm{core}}= 15.84(16)\times4\pi\epsilon_{0}a_{0}^{3}$  and $\alpha_{\textrm{vc}}=-0.67(20)\times4\pi\epsilon_{0}a_{0}^{3}$ \cite{Safronova2016}. It is important that these two values are computed by the same method for consistency. The uncertainty in the core contribution is taken to be 1\% based on the comparison with the coupled-cluster calculations \cite{Lim2002}. The uncertainty in the $\alpha_{\textrm{vc}}$ term is taken to be the difference of the RPA and Dirac-Hartree-Fock values.
%Calculations in the random-phase approximation yield for Cs, $\alpha_{\textrm{core}}= 15.84(32)\times4\pi\epsilon_{0}a_{0}^{3}$  and $\alpha_{\textrm{vc}}=-0.673\times4\pi\epsilon_{0}a_{0}^{3}$ \cite{Safronova2016}.  

The scalar polarizability contribution from the valence electrons is calculated by summing over contributions from all other states $\ket{k}=\ket{F',m_{F}'}$ \cite{LeKien2013}
\begin{equation}
\begin{split}
	\alpha_{\textrm{v},i}^{(0)}(\omega)&=\frac{2}{\hbar}\frac{\left(2F+1\right)}{\sqrt{3(2F+1)}}\sum\limits_{k\neq i}\frac{\omega_{k,i} |d_{k}|^2}{\omega_{k,i}^{2}-\omega^{2}}(-1)^{F'+F+1}\\
&\qquad\times
(2F'+1)
\begin{Bmatrix}
 F & 1 & F'  \\
   1 & F & 0 \\
\end{Bmatrix}
\begin{Bmatrix}
 F' & I & J' \\
   J & 1 & F \\
\end{Bmatrix}^{2},
\end{split}\label{eq:ScalarPolarizability}
\end{equation}
where $\omega_{k,i}=\omega_{k}-\omega_{i}$ is the transition frequency between $\ket{k}$ and $\ket{i}$, $\hbar$ is the reduced Planck constant, $d_{k}=\left <J||d||J'\right > $ is the reduced dipole matrix element between $\ket{i}$ and $\ket{k}$, and the terms in curly brackets are Wigner-6j symbols.  

\begin{figure}[bt!]
\includegraphics[scale =1.0]{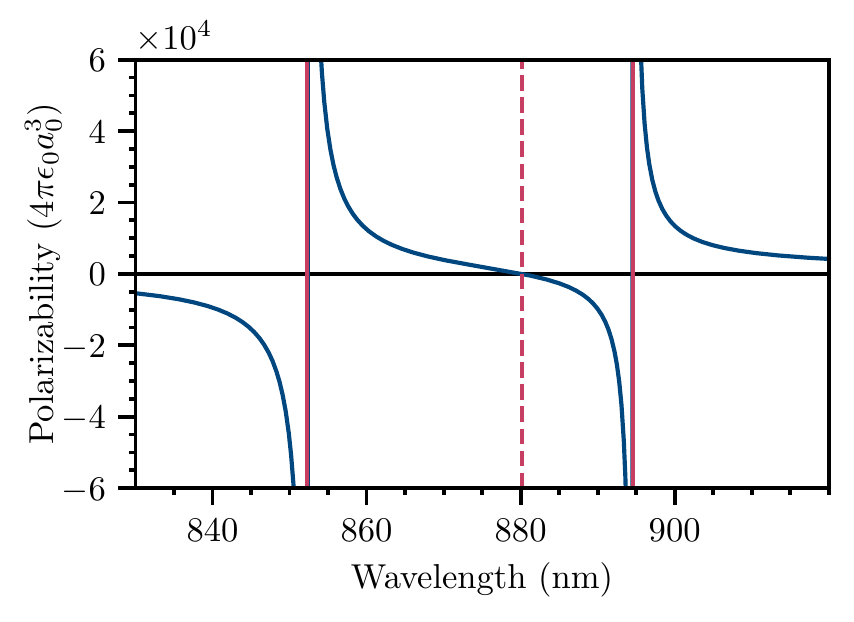}
\caption{Calculated scalar polarizability of ground state Cs in the vicinity of the $D_{1}$, $6S_{1/2}\rightarrow 6P_{1/2}$, transition at $894.6$ nm, and the $D_{2}$, $6S_{1/2}\rightarrow 6P_{3/2}$, transition at $852.3$ nm.  Between these two transitions the polarizability changes signs and passes through zero at the tune-out wavelength $\lambda_{0}\approx880$~nm as indicated by the dashed line. 
}
\label{fig:polarizability}
\end{figure}

It can be observed from Eq. \ref{eq:ScalarPolarizability}, that the scalar polarizability depends on the wavelengths of transitions from $\ket{i}$ that are allowed by electric dipole selection rules, and that the polarizability exhibits poles at these transition frequencies.  For ground state atoms, red (blue) detuned  frequencies lead to positive (negative) polarizability and hence attractive (repulsive) trapping potentials.  For atoms in the ground state, there are scalar tune-out wavelengths between all pairs of transitions at a wavelength where the red-detuned contribution to the polarizability from one transition is cancelled by the blue detuned contribution from the other transition.  Figure \ref{fig:polarizability} shows the calculated scalar polarizability of ground state Cs around the $D_{1}$ and $D_{2}$ transitions at $894.6$ nm and $852.3$ nm, respectively.  Between these two transitions, the polarizability goes to zero at $\lambda_{0}\approx880$~nm.  This is the tune-out wavelength that we measure in Sec. \ref{sec:Analysis}.

We now consider the impact of the vector and tensor polarizability terms  in Eq.~\ref{eq:svtPolarizability} on the value of the tune-out wavelength.  The vector polarizability
can cause substantial shifts to the tune-out wavelengths as a result of the selection rules for electric-dipole transitions.  To illustrate the importance of these selection rules, we consider Cs atoms in the $\ket{4,+4}$ ground state.  If the atoms interact with light polarized to drive $\sigma^{+}$ transitions ($|\xi|=1$), then transitions to the $6P_{1/2}$ state are not allowed by selection rules.  This lack of coupling to the $6P_{1/2}$ state means that no tune-out wavelength will be present between the $D_{1}$ and $D_{2}$ lines for this specific light polarization.  For the case studied here of atoms in $\ket{3, +3}$, all polarizations can couple to both the $6P_{3/2}$ and $6P_{1/2}$ states, but the position of the tune-out wavelength is still strongly influenced by the vector polarizability and can move on the order of $\sim10$\,nm for different polarizations.  We observe from Eq.~\ref{eq:svtPolarizability} that the vector polarizability contribution is proportional to the ellipticity of the light through the fourth Stokes parameter ($\xi$),  as well as the term $\hat{\mathbf{k}}\cdot\hat{\mathbf{B}}$.  We can therefore suppress the vector polarizability by ensuring the light polarization is highly linear and aligning the laser beam orthogonal to the magnetic field, so that ${\bf \hat{k}}\cdot{\bf \hat{B}}\rightarrow 0$.  Details of how this is achieved in our experiment are presented in Sec. \ref{sec:Measurements}.

The tensor polarizability term is relevant to the measurements performed here.  There is no contribution from the core electrons since the core is isotropic ($\alpha_{i}^{(2)}=\alpha^{(2)}_{v,i}$) \cite{Safronova2016}.  In the absence of hyperfine structure, the tensor polarizability is zero for the ground state.  However, including the hyperfine structure the tensor polarizability of $\ket{i}$ is non-zero,  and can also be written as a sum over states as \cite{LeKien2013}
\begin{equation}
\begin{split}
	\alpha_{\textrm{v},i}^{(2)}(\omega)&=\frac{2}{\hbar}\sqrt{\frac{10F(2F-1)(2F+1)}{3(F+1)(2F+3))}}\sum\limits_{k\neq i}\frac{\omega_{k,i}\left |d_{k} \right |^{2}}{\omega_{k,i}^{2}-\omega^{2}}\\
&\times
(-1)^{F+F'}(2F'+1)
\begin{Bmatrix}
 F & 1 & F'  \\
   1 & F & 2 \\
\end{Bmatrix}
\begin{Bmatrix}
 F' & I & J' \\
   J & 1 & F \\
\end{Bmatrix}^{2}.
\end{split}\label{eq:TensorPolarizability}
\end{equation}
For alkali-metal atoms in the ground state, the tensor term typically leads to corrections of less than a part-per-million.  Therefore, the polarizability for linearly polarized light is dominated by contributions from the scalar polarizability but small corrections due to the tensor polarizability need to also be taken into account.

\subsection{Kapitza-Dirac Scattering}

Kapitza-Dirac scattering~\cite{Kapitza1933} is routinely used in atomic physics experiments to measure optical lattice trap depths \cite{HeckerDenschlag2002,Viebahn2019, Gadway2009} and has previously been shown to be a useful tool for measuring tune-out wavelengths \cite{Schmidt2016, Kao2017}.  The technique has been extended to measure low lattice depths by applying multiple pulses of the lattice potential to the atoms \cite{Herold2012,Kao2017}. Here, we use Kapitza-Dirac scattering to measure the wavelength dependence of the atomic polarizability of Cs.
  
Kapitza-Dirac scattering occurs when the lattice is pulsed onto a Bose-Einstein condensate (BEC) and atoms in the condensate undergo stimulated two-photon scattering events.  Photons are scattered from one lattice beam to the other and therefore momentum transfer occurs in units of $2\hbar k_{\textrm{lat}}$, where $k_{\textrm{lat}}$ is the lattice wavevector.  The momentum transfer can occur in either direction along the beam to give both positive and negative momentum states.  As the lattice pulse time is varied the population will oscillate between the different $2l\hbar k_{\textrm{lat}}$ momentum states ($l$ is an integer).  The momentum states separate in a time-of-flight expansion allowing the populations to be measured.  In the Raman-Nath regime, the atomic motion during the lattice pulses can be neglected and analytic relations for the population dynamics can be used \cite{Gadway2009}.

In the work presented here, we consider Kapitza-Dirac scattering beyond the Raman-Nath regime \cite{Gadway2009}.  In this regime the pulses are no longer short compared to the oscillation period of atoms in the lattice.  We therefore cannot use analytic relations for the different momentum-state populations and a numerical model is required.  The Hamiltonian for atoms of mass $m$ in a periodic potential of depth $V_{0}$ and wavevector $k_{\textrm{lat}}$ applied in the $x$ direction is given by
\begin{equation}
    H=-\frac{\hbar^{2}}{2m}\frac{d^{2}}{dx^{2}}+V_{0}\sin^{2}(k_{\textrm{lat}}x).
\end{equation}
To calculate the populations in each momentum state, we diagonalize the Hamiltonian using a plane wave basis including both positive and negative orders up to $|l|=20$.  Convergence of solutions is found for $|l|>|l_{\textrm{max}}|$, where $|l_{\textrm{max}}|$ is the maximum populated momentum state.  In the measurements performed here, we observe populations in momentum states up to $|l_{\textrm{max}}|=5$. The atom numbers in each state are normalized by the total number of atoms in the image to avoid issues from shot-to-shot variations in the atom number.  The evolution of all momentum states are fit simultaneously, with the $\pm|l|2\hbar k_{\textrm{lat}}$ populations averaged during fitting to reflect symmetry of the scattering process. The only free parameters in the fit are the lattice depth and an amplitude factor to account for imperfect atom-number normalization.

\begin{figure}[bt!]
\includegraphics[scale=1]{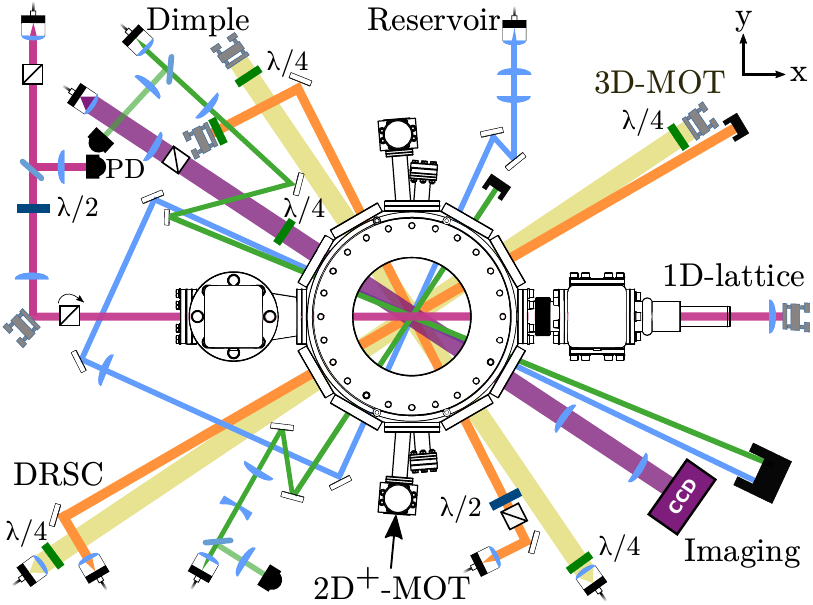}
\caption{
Schematic of the apparatus highlighting the optical beam layout in the x-y plane.  All beams used for cooling, trapping, and imaging of atoms are shown.  The 1D optical lattice used for  measuring the polarizability is aligned along the x-direction (left-right in the diagram).  Absorption imaging is performed at an angle of $33$\degree~with respect to the lattice direction.  Electromagnetic coils (not shown) above and below the chamber provide a magnetic bias field in the vertical direction.  The MOT and degenerate Raman sideband cooling beams in the $z$ direction are not shown for clarity. 
}
\label{fig:setup}
\end{figure}

\section{Overview of the apparatus and BEC production}\label{sec:Apparatus}

A schematic overview of our apparatus is shown in Fig.~\ref{fig:setup}.  Below we give brief details of the stages used to produce Cs BECs. 

Our experiment begins with a high-flux source of laser-cooled atoms from a 2D$^{+}$ magneto-optical trap (MOT) \cite{Dieckmann1998}.  Atoms from this source are collected in a 3D MOT in the center of a 12-port stainless steel chamber.
After sub-Doppler molasses cooling, degenerate Raman sideband cooling~\cite{Kerman2000} is performed which cools the atoms to $\sim1~\mu$K and polarizes them into the $\ket{F=3, m_{F}=+3}$ ground state.  To cool the atoms further we follow the method used to create the first BECs of Cs~\cite{Weber2003}.  In this approach, we implement a large volume reservoir trap consisting of two beams with waists of $\sim500~\mu$m at the atoms and crossing at an angle of 90 degrees.  The light for the reservoir trap is derived from a broadband Ytterbium fiber laser (IPG Photonics) with a wavelength around $1070$ nm.  The trap is setup in a bowtie configuration where the power is recycled and used in both of the trapping beams, as shown in Fig.~\ref{fig:setup}.  The reservoir trap requires a magnetic levitation gradient to support the atoms against gravity.

Approximately $10\%$ of the atoms are then transferred from the reservoir trap into a tighter crossed optical dipole trap (xODT) at $1064~$nm derived from a Nd:YAG laser (Coherent, Mephisto).  The two beams forming the trap have waists of $51(1)~\mu$m and $103(2)~\mu$m at the atoms.  Forced evaporation is then performed by reducing the powers of these xODT beams whilst applying a bias field of $23.4(1)~$G to minimise the 3-body inelastic loss rate~\cite{Weber2003}.  Typically, pure BECs containing $2\times10^{4}$ atoms in the $\ket{3,+3}$ ground hyperfine state are created.

\section{Polarizability Measurements}\label{sec:Measurements}

The lattice light used for the Kapitza-Dirac measurements is generated from a tuneable Ti:sapphire laser (M-Squared, SolsTiS) pumped by a $18$~W pump laser at $532$~nm (Lighthouse Photonics, Sprout).  The light intensity sent to the experiment is controlled by an acousto-optical modulator (AOM) that uses a fast switch (Mini-Circuits, ZASWA2-50DR-FA+) to generate the short pulses required for the measurements.  The light from the AOM is coupled into an optical fiber to avoid changes in the lattice alignment as the wavelength of the laser is adjusted.  The power output of the fiber is monitored using a photodiode as shown in Fig. \ref{fig:setup}.  This photodiode is used to correct for small power changes between polarizability measurements.  
Before passing through the vacuum chamber, the lattice light passes through a Glan-Laser polarizer (Thorlabs, GL10-B).  This polarizer minimizes $\xi$ and achieves a highly polarized lattice beam which is linearly polarized with an extinction ratio of better than $10^{-5}$.
The waist of the ingoing lattice beam is measured to be $99(5)~\mu$m at the position of the atoms.  After the light has passed through the chamber, it is collimated and retro-reflected onto the atoms to create the lattice potential.  The lattice laser frequency is measured and stabilised using a HighFinesse \mbox{WS-U} wavemeter with an absolute accuracy  of $30$ MHz.  We reference the wavemeter to a laser frequency stabilised to the $5^{2}S_{1/2} (F=2)\rightarrow 5^{2}P_{3/2} (F'=3)$ transition in $^{87}$Rb.

\begin{figure}[bt!]
\includegraphics{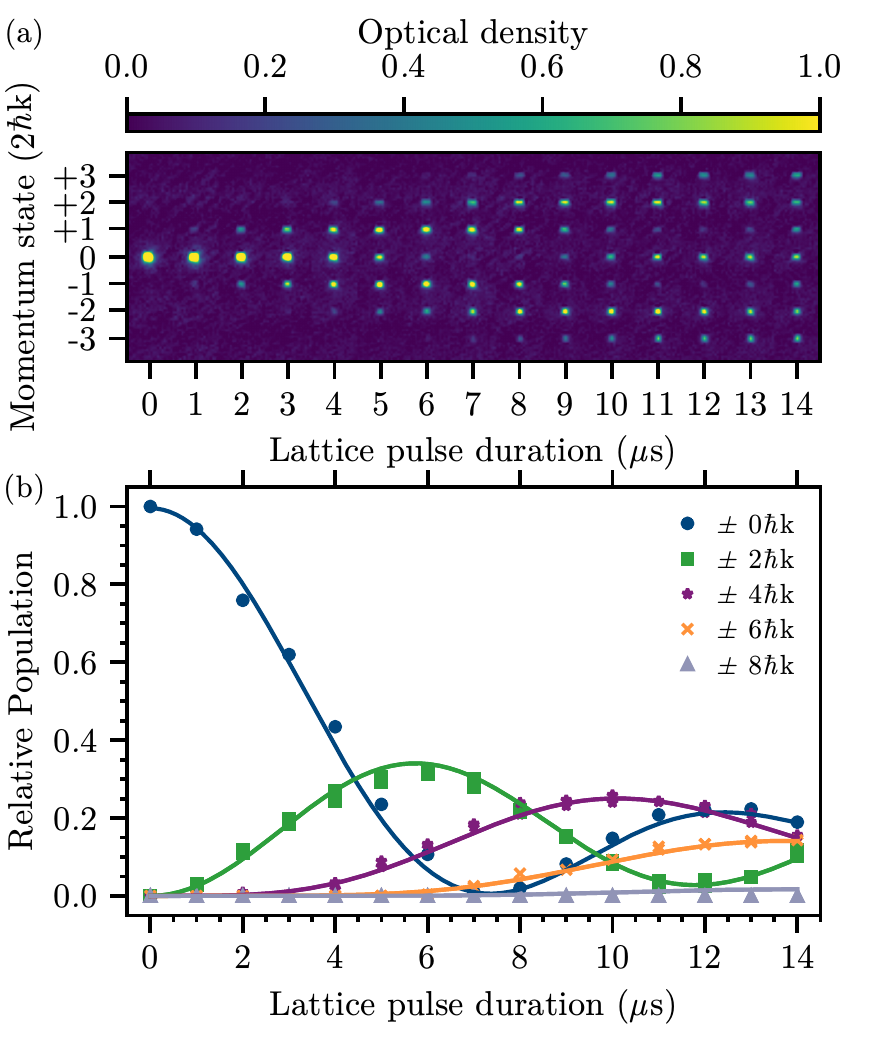}
\caption{ 
An example of a Kapitza-Dirac measurement for a lattice of power $P\sim300$ mW at a wavelength of $881$ nm.  (a) Absorption images of the different BEC momentum states for varying lattice pulse duration, measured after a $40$\,ms levitated time-of-flight.  (b)  The relative atom number of the different diffracted momentum states are extracted from the images and fit using the method described in the text to give a trap depth of $4.96(2)~\mu$K.
} 
\label{fig:KDpopulation evolution}
\end{figure}

To perform the measurements of the polarizability, the BEC is released from the dipole trap and, after a 100~$\mu$s delay, the lattice is pulsed on for a variable time.  The atoms are then levitated for $40$~ms using a magnetic field gradient of $\approx30$~G/cm, allowing the different momentum peaks to separate spatially before being imaged.  Example images from such diffraction measurements are shown in Fig.~\ref{fig:KDpopulation evolution}(a) for a lattice created using $\sim300$~mW of $881$~nm light and applied for varying pulse duration.  Figure~\ref{fig:KDpopulation evolution}(b) shows the extracted populations of each of the momentum states for each of these images.  

Birefringence in the viewports of the vacuum chamber can cause the highly linearly polarized light to  acquire a circularly polarized component.  In order to suppress any vector polarizability contribution caused by the vacuum viewports, we therefore perform separate measurements using two orthogonal linear polarizations and then average the two measured tune-out wavelengths~\cite{Herold2012}. We choose the two lattice polarization alignments to be parallel and orthogonal to the applied magnetic field.  This choice of orthogonal polarizations has the advantage that the changes in tune-out wavelength due to the tensor polarizability are less sensitive to alignment of the polarization in these orientations. 

\begin{figure*}[hbt!]
\includegraphics{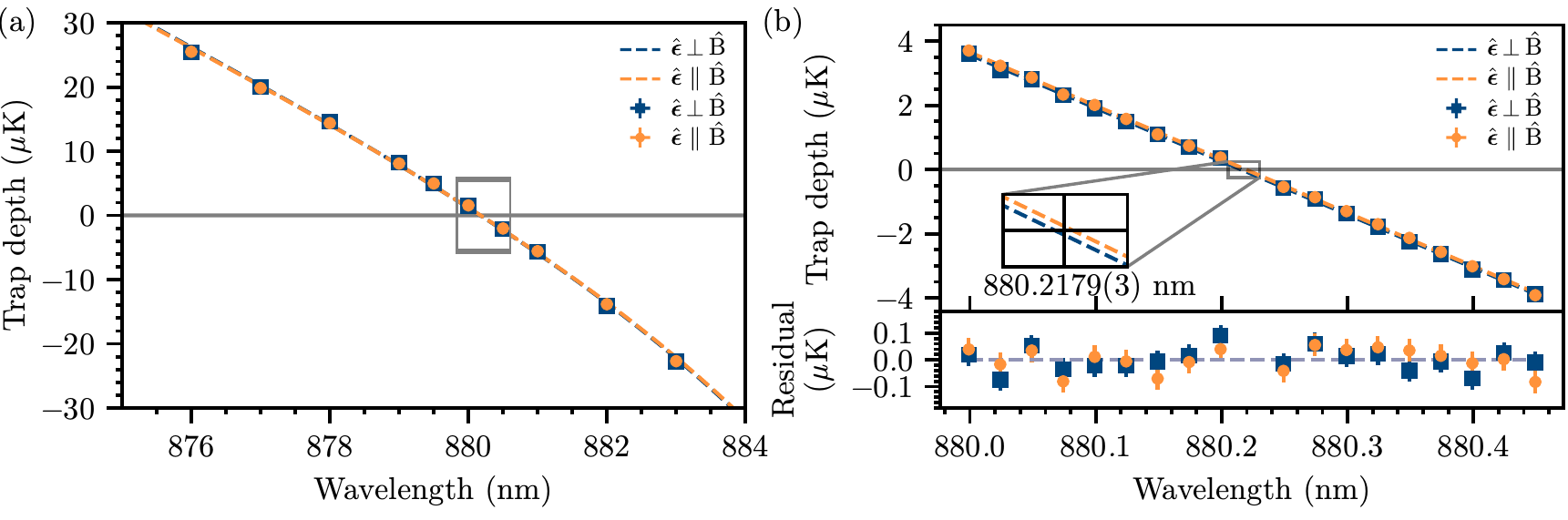}
\caption{Trap depth measurements using Kapitza-Dirac scattering for polarization vector, $\hat{\boldsymbol{\epsilon}}$, parallel ($\parallel$, orange, circles) and perpendicular ($\perp$, blue, squares) to the magnetic field $\hat{\boldsymbol{B}}$. (a) A broad wavelength scan of trap depth using $P\sim300$ mW. Dashed lines are fitted trap depths (see text for details).  (b)  Narrow wavelength scan across the range of wavelengths shown by the grey box in (a).  Trap depths are measured using a lattice power of $P\sim 1$~W, and are fitted with straight line functions to extract tune-out wavelengths $\lambda_{0}$ for each polarization.  The residuals are plotted in the lower panel.  The fitted tune-out wavelengths are $\lambda_{0}^{\perp}=880.2164(6)$~nm and $\lambda_{0}^{_{\parallel}}=880.2195(6)$~nm for lattice polarizations perpendicular and parallel to $\hat{\boldsymbol{B}}$, respectively.  The difference in these values is highlighted in the inset of (b).
}
\label{fig:Results}
\end{figure*}

We initially measure the trap depth of the lattice using a power of $\sim300$~mW.  
We measure trap depths for different lattice wavelengths and different orientations of the lattice polarization with respect to the magnetic field.  Figure~\ref{fig:Results}(a) shows the results of these measurements together with a fit using Eq.~\ref{eq:TotalScalar}.  In the fit, the weighting of the polarizability contributions from the $D_{1}$ and $D_{2}$ transitions is a free parameter and the polarizability contributions from the other transitions is assumed to be wavelength independent over this range.  The overall amplitude of the fit is also a fit parameter to convert from polarizability to trap depth.

To measure the tune-out wavelength, we increase the power to $\sim1$~W, increasing our sensitivity to small polarizabilities and allowing measurements to be made closer to $\lambda_0$. We then performed measurements of the trap depth over a $\sim0.5$~nm wavelength range centred on the tune-out wavelength, as shown in Fig.~\ref{fig:Results}(b).
The polarizability can be extracted from the trap depth if the powers, beam waists and beam overlap are all known.  However, to determine the tune-out wavelength only relative changes in polarizability are required if the lattice beam parameters remain constant.  We therefore use the extracted trap depth to determine the tune-out wavelength.

\section{Analysis of results}\label{sec:Analysis}

In order to extract the tune-out wavelength, from the measurements shown in Fig.~\ref{fig:Results}(b), we first fit the data using a linear function which is a good approximation to the polarizability over this wavelength range.  We separately fit the data for the orthogonal linear polarizations.  The fits yield the tune-out wavelengths of $\lambda_{0}^{\perp}=880.2164(6)$~nm and $\lambda_{0}^{_{\parallel}}=880.2195(6)$~nm, where $\perp$ ($\parallel$) indicates that the linear polarization of the lattice is orthogonal (parallel) to the magnetic field.  To extract the scalar tune-out wavelength from these measurements we rely on some theoretical corrections detailed below.  

\subsection{Corrections to Measurements}

The first correction that is taken into account is to remove the shifts from the tensor polarizability.  From Eq.~\ref{eq:svtPolarizability} we see that the shift depends on the relative orientation of the magnetic field and the polarization of the lattice.  Using Eq.~\ref{eq:TensorPolarizability}, we determine the tensor shifts to be $473$~fm when $\hat{\boldsymbol{\epsilon}}\cdot\hat{\mathbf{B}}=1$ and $-237$~fm when  $\hat{\boldsymbol{\epsilon}}\cdot\hat{\mathbf{B}}=0$.  We note that the uncertainties in these calculated values are irrelevant compared to our statistical uncertainty in the measurement of $\lambda_0$. 

After applying this correction the tune-out wavelengths become $\lambda_{0}^{\perp}=880.2166(6)$~nm and $\lambda_{0}^{_{\parallel}}=880.2190(6)$~nm. Although the two values agree at the $2\sigma$ level, the small difference between them may also indicate the presence of a residual vector polarizability shift due to the birefringence in the vacuum viewports. We cancel this effect by averaging the tune-out wavelength measurements from the two orthogonal polarizations, giving the result $\lambda_{0}=880.2178(4)$~nm.

%The remaining difference between these two values comes from two factors:  The first of these differences is statistical uncertainty where the two values agree at the $2\sigma$ level. The second difference can also come from any remaining vector polarizability.  Any vector polarizability shift coming from the viewports can be cancelled by averaging the tune-out wavelength measurements from the two orthogonal polarizations, giving the result $\lambda_{0}=880.2178(4)$~nm.

The next correction we make is for the applied magnetic bias field.  We calibrate the magnetic field in our experiment using the known Cs Feshbach resonances up to $50$~G \cite{Chin2004}.  We wish to extrapolate our measurements to the case of zero applied magnetic field.  The shift is calculated following the method given in \cite{Leonard2015}.  We consider only the transitions to the $6P_{J}$ states and calculate the shift to be $65(3)$~fm.  Including transitions to the $7P_{J}$ states results in shifts of $\sim0.002$~fm.  Applying this correction gives the frequency-dependent scalar tune-out wavelength to be $\lambda_{0}=880.2179(4)$~nm, where the error bar is purely statistical. 

We must also include the systematic errors in our final result.  The first systematic error comes from our wavelength measurement.  The wavemeter has an uncertainty of $30$~MHz corresponding to $78$~fm at this wavelength.  The other major error comes from the vector polarizability  which we calculate to have a conservative upper bound of $24$~fm assuming we have aligned the lattice within 10 degrees of being orthogonal to the  magnetic field quantization axis.  This gives us the final result for the tune-out wavelength $\lambda_{0}=880.21790(40)_{\textrm{stat}}(8)_{\textrm{sys}}$~nm where we separate the statistical and systematic uncertainties. 

There are a number of theoretical predictions for $\lambda_{0}$ \cite{Arora2011,Yu2016,Jiang2020a}.  However, only one of these theoretical values takes into account the hyperfine structure of the atoms \cite{Jiang2020a} and gives a predicted tune-out wavelength of $880.20(5)~\text{nm}$.  Our measured tune-out wavelength agrees well with this result.  

\begin{table}
\caption{The theoretical calculations of the $6s-6p_{1/2}$ and $6s-6p_{3/2}$ matrix elements  and their ratio  using   four variants of the all-order method; \textit{ab initio} linearized coupled-cluster results with single-double (SD) and perturbative valence triple (SDpT) excitations are given in the SD and SDpT columns, and  scaled values are listed in the SDsc and SDpTsc columns. }
\begin{ruledtabular}
\begin{tabular}{l cccc}% lcr
	&	SD & SDsc  &	SDpT & SDpTsc\\
\hline
$6S-6P_{1/2}$&	4.4807	&4.5350&	4.5576&	4.5302\\
$6S-6P_{3/2}$&	6.3030	&6.3818&	6.4136&	6.3734\\
Ratio	     &   1.9788	&1.9803&	1.9803&	1.9793
\end{tabular}
\end{ruledtabular}\label{tab:ratio}
\end{table}

\subsection{Ratio of Matrix Elements}
The error in the theoretical value of the tune-out wavelength is dominated by the ratio of the $6P$ matrix elements.  It is therefore interesting to use our measurements to extract a value for this ratio, which is defined as
\begin{equation}
    R=\frac{\left |\left <6P_{3/2}||d||6S_{1/2}\right > \right |^{2}}{\left |\left <6P_{1/2}||d||6S_{1/2}\right > \right |^{2}}=\frac{|d_{6P_{3/2}}|^{2}}{|d_{6P_{1/2}}|^{2}}\quad .
\end{equation} 
When considering degeneracies of states only, this ratio is expected to be $R=2$.  However, including relativistic corrections, which are large for Cs compared to other alkali-metal atoms due to the large atomic mass, this ratio is reduced, with a theoretical value of $R=1.984(10)$ \cite{Jiang2020a}.

We calculate the ratio $R$ using the relativistic linearized coupled-cluster method \cite{Safronova2016}. The results of four computations are listed in Table \ref{tab:ratio}. \textit{Ab initio} linearized coupled-cluster results with single-double (SD) and perturbative valence triple (SDpT) excitations are given in the SD and SDpT columns, and the scaled values are listed in SDsc and SDpTsc columns. These approaches are described in \cite{Safronova2008}. A large fraction of the correlation correction cancels for the ratio, and its accuracy is substantially higher than that of the matrix elements.  The final value is taken to be 1.9788(21), in excellent agreement with the experiment. All three approximations beyond SD are aimed at evaluating one type of the correlation corrections (the so called ``Brueckner-orbital (BO) correction''), and its uncertainty is evaluated as the spread of the four results. The total of all other corrections is of the same order as the BO correction and we assume their total uncertainty to be similar to the uncertainty of BO correction, based on comparison with lifetime measurements. The present work validates this procedure for the ratio as well.

Table \ref{tab:contributions} shows the contributions to the frequency dependent scalar polarizability at the theoretical tune-out wavelength, between the $D$ line transitions, of $\lambda_{0}^{\text{th}}=880.2463$~nm.  To determine $\lambda_{0}^{\text{th}}$, we use the matrix elements given in Table \ref{tab:contributions} which come from a mixture of both experimentally measured and theoretically calculated values.  From the table it can be seen that the main contributions to the polarizability at this tune-out wavelength are from the transitions to the $6P_{1/2}$ and $6P_{3/2}$ states, with the other values constant around this value.  Therefore by adjusting the ratio $R$ the theoretical value of the tune-out wavelength can be adjusted to agree with the measured value.

In order to determine this ratio of matrix elements, the scalar polarizability can be expressed in the following form~\cite{Leonard2015}
\begin{equation}
    \alpha_{6S_{1/2}}^{(0)}(\omega)=\alpha_{\textrm{offset}} + \left|d_{6P_{1/2}}\right|^{2}\left(K_{6P_{1/2}}+K_{6P_{3/2}}R\right),\label{eq:fitfunc}
\end{equation}
\noindent where $\alpha_{\textrm{offset}}=17.5(3)\times 4\pi\epsilon_{0}a_{0}^{3}$ includes all contributions to the scalar polarizability that are not from the $6P_{1/2}$ and $6P_{3/2}$ states. $K_{6P_{J}}=\alpha^{(0)}_{6P_{J}}/|d_{6P_{J}}|^{2}$ where $\alpha^{(0)}_{6P_{J}}$ are the polarizability contributions to the $6S$ state from the $6P_{J}$ states.

Setting $\alpha=0$ in Eq. \ref{eq:fitfunc} and using our experimentally measured value for the tune-out wavelength we extract the ratio $R=1.9808(2)$.  The uncertainty in $R$ contains contributions from $d_{6P_{1/2}}$, $\alpha_{\textrm{offset}}$ and the determination of the tune-out wavelength.  However, the dominant contribution comes from the calculation of the $\alpha_{\text{core}}$.  Using the experimentally measured values of the $6P_{J}$ states given in Table \ref{tab:contributions}, the value of the ratio is $R=1.984(4)$ which agrees well with our value extracted from the measured tune-out wavelength.
%\sarah{Paragraph here on a full calculation of ratio.}

\begin{table}
\caption{The theoretical contributions to the 6$S$ scalar polarizability of Cs at $\lambda_{0}^{th}$ = 880.2463 nm.  Polarizability contributions are given in units of ($4\pi \epsilon_{0}a_0^3$). Uncertainties are given in parentheses. Experimental energies $\Delta E$ are measured from the ground state and given in cm$^{-1}$ \cite{Sansonetti2009}. The reduced electric-dipole matrix elements $d$ in atomic units are from experimental and theoretical data.  }
\begin{ruledtabular}
\begin{tabular}{l l l l l}% lcr
State	&	$\Delta E~(\text{cm}^{-1}) $ & $d$ (ea$_0$)  &	$\alpha^{(0)}$ (at $\lambda_{0}^{\text{th}}) $& Ref.\\   
\hline
$6P_{1/2} $ & 11178.26816  	&  4.489(6)     & -4029(11)     & \cite{Rafac1999}\\
$7P_{1/2} $ & 21765.348	   	& 0.2781(5)     & 0.3573(12)    & \cite{Damitz2019}\\
$8P_{1/2} $ & 25708.8547	& 0.081(3)     & 0.030(7)      &\cite{Safronova2016} \\
$9P_{1/2} $ & 27636.9966	& 0.043(7)      & 0.006(2)      &\cite{Safronova2016} \\
$10P_{1/2} $ & 28726.8123	& 0.0248(5)     & 0.0019(8)     &\cite{Safronova2016} \\
$11P_{1/2} $ & 29403.42310	& 0.0162(4)     & 0.0008(4)     &\cite{Safronova2016} \\
$12P_{1/2} $ & 29852.43153	& 0.012(3)      & 0.0004(2)     & \cite{Safronova2016}\\
$6P_{3/2} $ & 11732.3071    & 6.335(5)      & 4011(6)       &\cite{Patterson2015}	\\
$7P_{3/2} $ & 21946.397	    & 0.5742(6)     & 1.501(3)      &\cite{Damitz2019}	\\
$8P_{3/2} $ & 25791.508	  	& 0.232(14)     & 0.19(2)	    &\cite{Safronova2016}\\
$9P_{3/2} $ &  27681.6782	& 0.130(10)     & 0.053(8)      &\cite{Safronova2016} \\ 
$10P_{3/2} $ & 	28753.6769  & 0.086(7)      & 0.022(4)      &\cite{Safronova2016}	\\
$11P_{3/2} $ &  29420.824	& 0.063(6)      & 0.011(2)      &\cite{Safronova2016} \\ 
$12P_{3/2} $ &  29864.345	& 0.049(5)      & 0.0068(13)      & \cite{Safronova2016}\\
$n>12$ &  &               & %2.18(2)+\sarah{0.16(16)}
0.16(16)
& 
%\cite{Safronova2016}
\\
$\alpha_{\text{core}}$ & &  &   15.84(16)& \cite{Safronova2016}\\
$\alpha_{\text{vc}}$ && &  -0.67(20) &\cite{Safronova2016} \\
$\alpha_{\text{offset}}$ &  &&
%17.3(3)+\sarah{tail 0.16(16)} 
17.5(3)
&  \\
Total & &  &0(13) &   \\
\end{tabular}
\end{ruledtabular}\label{tab:contributions}
\end{table}

\section{Summary and Outlook}\label{sec:Outlook}

We have used Kapitza-Dirac scattering of atoms from a 1D optical lattice to measure the tune-out wavelength of Cs in the ground hyperfine state,
$\ket{F=3, m_{F}=3}$, around $880~\text{nm}$ between the $D_{1}$ and $D_{2}$ transitions.  We are able to eliminate the influence of the vector Stark shift by using linearly polarized light and performing the measurement for two orthogonal polarizations.  By correcting the measured wavelength to remove the influence of the tensor polarizability and magnetic field, we find the scalar tune-out wavelength to be $880.21790(40)_{\text{stat}}(8)_{\text{sys}}$~nm.  This is in good agreement with theoretical value of $880.20(5)~\text{nm}$ from \cite{Jiang2020a}.

We have used this measurement to determine the ratio of the reduced matrix elements for transitions from the ground state to the $6P_{J}$ states.  We have found this ratio to be $R=1.9808(2)$ which is consistent with previous measurements, but with a reduction in the uncertainty by a factor of more than $10$.  This ratio of reduced matrix elements is also important for determining other atomic properties, such as oscillator strengths and state lifetimes. The present  work provides a benchmark test of the relativistic all-order method and the procedure used to evaluate the uncertainty in the theory. This is particularly important, as this method is used to generate data and the associated uncertainties for the online data portal \cite{UDportal}.

The measurement of this tune-out wavelength will be useful in future studies of quantum degenerate mixtures involving Cs~\cite{Lercher2011,McCarron2014,Grobner2016,Wilson2021,Warner2021}. For example, a stirring beam at this wavelength could be used to create vortices in the atomic species trapped with Cs, without affecting the Cs condensate.  In future work we plan to measure the tune-out wavelengths for Cs in the vicinity of the $6^2S_{1/2}\rightarrow7^2P_{1/2},7^2P_{3/2}$ transitions to put constraints on additional dipole matrix elements.  Such measurements are also more sensitive to changes in $\alpha_{\text{core}}$~\cite{Holmgren2012} and may contribute to further constraints on this parameter.

\section{Acknowledgements}
This work was supported by U.K. Engineering and Physical Sciences Research Council (EPSRC) Grants EP/P01058X/1 and EP/N007085/1.  SLB acknowledges support from the EU Marie Sk\l odowska-Curie program [Grant No. MSCA-IF-2018-838454]. M.S.S. acknowledges the sponsorship of US ONR Grant No. 
N00014-20-1-2513. The data, code and analysis associated with this work are available at: DOI to be added.

%\bibliography{TuneOutBib}
%

\end{document}